\documentclass[aps,prb,reprint,groupedaddress,showpacs,footinbib,citeautoscript]{revtex4-1}

\usepackage{graphicx}
\usepackage{amsmath}
\usepackage{mathrsfs}
\usepackage{amssymb}
\usepackage{bm}
\usepackage{subfigure}
\usepackage{array}

\usepackage{float}
\usepackage{color}

\graphicspath{{.}{./EPS/}}

\newcommand{\ket}[1]{\left | \, #1 \right \rangle}

\newcommand* {\ee}{\ensuremath{\mathrm{e}}}
\newcommand*{\vek}[1]{{\ensuremath{\bm{\mathrm{#1}}}}}

\newcommand*{\kk}{{\bm{\mathrm{k}}}}
\newcommand*{\rr}{{\bm{\mathrm{r}}}}
\newcommand*{\qq}{{\bm{\mathrm{q}}}}
\newcommand*{\pp}{{\bm{\mathrm{p}}}}

\usepackage[colorlinks=true,citecolor=blue]{hyperref}
\hypersetup{colorlinks=true,citecolor=blue,linkcolor=red,urlcolor=blue}

\begin{document}

% Use the \preprint command to place your local institutional report
% number in the upper righthand corner of the title page in preprint mode.
% Multiple \preprint commands are allowed.
% Use the 'preprintnumbers' class option to override journal defaults
% to display numbers if necessary
%\preprint{}

%Title of paper
\title{Spin susceptibility of two-dimensional transition metal dichalcogenides}

% repeat the \author .. \affiliation  etc. as needed
% \email, \thanks, \homepage, \altaffiliation all apply to the current
% author. Explanatory text should go in the []'s, actual e-mail
% address or url should go in the {}'s for \email and \homepage.
% Please use the appropriate macro for each type of information

% \affiliation command applies to all authors since the last
% \affiliation command. The \affiliation command should follow the
% other information
% \affiliation can be followed by \email, \homepage, \thanks as well.

\author{H. Hatami}
\affiliation{School of Chemical and Physical Sciences and MacDiarmid Institute
for Advanced Materials and Nanotechnology, Victoria University of Wellington,
PO Box 600, Wellington 6140, New Zealand}

\author{T. Kernreiter}
\affiliation{School of Chemical and Physical Sciences and MacDiarmid Institute
for Advanced Materials and Nanotechnology, Victoria University of Wellington,
PO Box 600, Wellington 6140, New Zealand}

\author{U. Z\"ulicke}
\email{uli.zuelicke@vuw.ac.nz}
\affiliation{School of Chemical and Physical Sciences and MacDiarmid Institute
for Advanced Materials and Nanotechnology, Victoria University of Wellington,
PO Box 600, Wellington 6140, New Zealand}

\date{\today}

\begin{abstract}
We have obtained analytical expressions for the $q$-dependent static spin
susceptibility of monolayer transition metal dichalcogenides, considering
both the electron-doped and hole-doped cases. Our results are applied to
calculate spin-related physical observables of monolayer MoS$_2$,
focusing especially on in-plane/out-of-plane anisotropies. We find that
the hole-mediated RKKY exchange interaction for in-plane impurity-spin
components decays with the power law $R^{-5/2}$ as a function of
distance $R$, which deviates from the $R^{-2}$ power law normally
exhibited by a two-dimensional Fermi liquid. In contrast, the out-of-plane
spin response shows the familiar $R^{-2}$ long-range behavior. We also
use the spin susceptibility to define a collective $g$-factor for hole-doped
MoS$_2$ systems and discuss its density-dependent anisotropy. 
\end{abstract}

% insert suggested PACS numbers in braces on next line

\maketitle

%%%%%%%%%%%%%%%%%%%%%%%%%%%%%%%%%%%%%%%%%%%%%%%
%%%%%%%%%%%%%%%        INTRODUCTION      %%%%%%%%%%%%%%%%%%%
%%%%%%%%%%%%%%%%%%%%%%%%%%%%%%%%%%%%%%%%%%%%%%%
\section{Introduction}
% Put \label in argument of \section for cross-referencing
%\section{\label{}}

The discovery of graphene\cite{nov04,cas09}, a monolayer of carbon atoms arranged in a
honeycomb lattice, and its intriguing physical properties has triggered a search for other
materials that, like graphene, are intrinsically two-dimensional (2D). Despite its huge potential
for applications in electronic devices\cite{bol08,mor08,han07,nov12}, there are seveal reasons
to consider alternatives to graphene. An important motivation is provided by the fact that
pristine graphene is a semimetal, i.e., its conduction and valence bands touch at the neutrality
(Dirac) point. The absence of an energy gap creates difficulties for realizing graphene-based
conventional semiconductor devices\cite{and09}. Furthermore, graphene has a very weak
spin-orbit coupling (SOC)\cite{min06,kon10}. Having a similar material with strong intrinsic
SOC would open up possibilities for pursuing novel (e.g., magnet-less) spintronic
applications\cite{zut04}.

Two-dimensional crystals of transition metal dichalcogenides have recently been identified
as graphene-like materials that have very interesting properties and great promise for
enabling electronic and spintronic applications\cite{wan12,xu14}. As a member of this materials
class, monolayer MoS$_2$ has attracted a lot of attention recently\cite{rad11,zen12,cao12}.
While bulk MoS$_2$ is an indirect-gap semiconductor, a monolayer is found to be
semiconducting with a direct band gap\cite{mak10,xia12,kor13,ros13,cap13,che13a}.
Monolayer MoS$_2$ has a honeycomb-lattice structure with Mo and S atoms located on
different sublattices. This arrangement gives rise to a broken inversion symmetry, which in
turn yields a relatively large band gap ($\sim\,$1.66$\,$eV). Due to the relevant admixture of
Mo $d$-orbitals, monolayer MoS$_2$ also has a strong SOC, rendering it a good candidate
for spintronic applications\cite{nea13,kli13}. Recent theoretical studies of MoS$_2$ have
focused on the many-particle and collective response properties of its charge carriers.
Plasmon dispersions and static screening have been investigated within the random
phase approximation\cite{sch13}. Other works\cite{lu13,son13,och13,och13a,wan13,yu14}
have discussed the various spin-relaxation processes that can occur in MoS$_2$.
Furthermore, the carrier-mediated exchange interaction between localized magnetic
impurities has been calculated\cite{par13} within the framework of the RKKY
mechanism\cite{rud54,kas56,yos57} and using first-principle methods~\cite{che13,mis13}.
A recent study~\cite{dol13} has systematically explored realistic strategies for achieving
\textit{n}-type and \textit{p}-type doping in monolayer MoS$_2$.

Our work sheds new light on the spin response of 2D transition metal dichalcogenides.
Analytical results for the wave-vector-dependent spin susceptibility\cite{mor85,yos96}
$\chi_{ij}(\qq)$ are obtained based on $\kk\cdot\pp$ model-Hamiltonian
descriptions\cite{xia12,kor13,ros13}. Physical consequences are discussed and illustrated
using band-structure parameters for MoS$_2$. We reveal interesting features exhibited by
carrier-mediated exchange interactions between local magnetic moments and Zeeman
spin splitting as encoded in the electronic $g$-factor. The
hole-doped material turns out to have particularly rich spin properties, whereas the
electron-doped case shows behavior quite similar to that of ordinary 2D electron systems.
Nevertheless, from a conceptual point of view, consideration of the electron-doped material
is useful because it serves as an instructive testbed for understanding the interplay between
extrinsic and intrinsic contributions to the spin response, where the former (latter) result
from filled states in the conduction (valence) band.
Thus the spin-response properties of monolayer transition metal dichalcogenides constitute
an intriguing intermediate behavior between that exhibited by graphene and ordinary 2D
electron systems realized in semiconductor heterostructures. Besides adding to the basic
understanding of a new materials class, our results also suggest practical ways for
electronic manipulation of its spin structure.

The remainder of this article is organized as follows. In Sec.~\ref{sec:model}, we introduce
the low-energy effective Hamiltonian that underpins our calculations and present
a general expression for the spin-susceptibility tensor in terms of single-particle eigenstates
and -energies. Analytical results for $\chi_{ij}(\qq)$ are presented in the subsequent two
sections; for the electron-doped case in Sec.~\ref{sec:susElec} and for the hole-doped
case in Sec.~\ref{sec:susHole}. We discuss salient features of spin-related physical
observables for the hole-doped system in Sec~\ref{sec:obs}. Section~\ref{sec:sum} contains
a summary of our results and conclusions. A detailed derivation of the basic expression
for the spin susceptibility used in our work is provided in the Appendix.

%%%%%%%%%%%%%%%%%%%%%%%%%%%%%%%%%%%%%%%%%%%%%%%
%%%%%%%%%%%%%%%               MODEL              %%%%%%%%%%%%%%%%%%%
%%%%%%%%%%%%%%%%%%%%%%%%%%%%%%%%%%%%%%%%%%%%%%%
\section{Details of theoretical approach\label{sec:model}}

\subsection{Model-Hamiltonian description}

As our formal basis for the calculation of the spin susceptibility, we adopt the low-energy
effective Hamiltonian for monolayer transition metal dichalcogenides derived in
Ref.~\onlinecite{xia12} (see also Refs.~\onlinecite{kor13,ros13}). To lowest order in the
in-plane wave vector $\kk=(k_x, k_y)$, it reads~\footnote{We neglect the recently
discussed~\cite{kor13,ros13} corrections to effective band masses and trigonal warping,
which only give rise to small quantitative corrections to the spin susceptibility.}
\begin{equation}\label{eq:MoS2Ham}
H^\tau_0 = a t(\tau k_{x}{\hat \sigma_{x}} +k_{y}\hat \sigma_{y})\otimes\openone
+\frac{\Delta}{2}\hat \sigma_{z}\otimes\openone - \frac{\lambda\tau}{2}\left(\hat \sigma_{z}
-\openone\right) \otimes \hat s_{z}~.
\end{equation}
The valley index $\tau=\pm1$ distinguishes electronic excitations at the two nonequivalent
high-symmetry points $\vek{K}$ and $\vek{K'}\equiv -\vek{K}$ in the Brillouin zone. The
symbol $a$ denotes the lattice constant, $t$ is the nearest-neighbor hopping matrix
element, $\Delta$ is the fundamental energy gap between conduction and valence bands,
and $2\lambda$ is a measure of the material's intrinsic spin-orbit coupling strength. The
Pauli matrices $\hat\sigma_{x,y,z}$ act in the space of basis functions for the conduction
and valence-band states at the $\vek{K}$ and $\vek{K'}$ points. In contrast, $\hat s_z$ is
the diagonal Pauli matrix associated with the charge carriers' real spin. For the case of
MoS$_2$, values of the relevant parameters are\cite{xia12} $a=3.193$~\AA, $t=1.1$~eV,
$\Delta=1.66$~eV, and $2\lambda=0.15$~eV. These values have been used in our
calculations whose results are plotted in figures.

The term proportional to $\lambda$ in Eq.~(\ref{eq:MoS2Ham}) breaks the spin-rotational
invariance in our system of interest; with eigenstates having their real spin quantized
along the out-of-plane ($z$) direction. In the following, we use a representation where the
space of conduction (c) and valence (v) bands is combined with the real-spin space, and
we will adopt the states $\ket{\mathrm{c}\!\uparrow}$,  $\ket{\mathrm{v}\!\uparrow}$,
$\ket{\mathrm{c}\!\downarrow}$, $\ket{\mathrm{v}\!\downarrow}$ from each individual
valley as our basis. The generalized Pauli matrices for real spin are then given by
$\hat{\mathcal J}_i=\hat J_i\otimes\hat\tau_0$, with $\hat\tau_0\equiv\openone_{2\times 2}$
being the identity matrix in valley space, and
\begin{subequations}\label{eq:SpinMat}
\begin{equation}
\hat J_x=\begin{pmatrix} 0 & 0 &1&0\\
0 & 0 &0&1\\
1 & 0 &0&0  \\
0 & 1 &0&0
\end{pmatrix}~,
\end{equation}
\begin{equation}
\hat J_y=\begin{pmatrix} 0 & 0 &-i&0\\
0 & 0 &0&-i\\
i & 0 &0&0 \\
0 & i & 0&0
\end{pmatrix}~,
\end{equation}
\begin{equation}
\hat J_z=\begin{pmatrix} 1 & 0 &0&0\\
0 & 1 &0&0\\
0 & 0 &-1&0  \\
0 & 0 &0&-1
\end{pmatrix}~.
\end{equation}
\end{subequations}
It is instructive to express also the model Hamiltonian given in Eq.~(\ref{eq:MoS2Ham})
as a matrix corresponding to our chosen representation. Using polar coordinates $\kk =
(k, \theta)$ for the in-plane wave vector, we find
\begin{equation}\label{eq:MoS2HamMat}
H_0^\tau = \begin{pmatrix} \frac{\Delta}{2} &at \tau k \ee^{-i \tau\theta}   &0&0 \\[2mm]
a t \tau k \ee^{i\tau\theta} &- \frac{\Delta}{2}+ \lambda\tau&0&0\\[2mm]
0&0&\frac{\Delta}{2} &a t\tau k \ee^{-i\tau\theta}  \\[2mm]
0&0&a t\tau k \ee^{i \tau\theta}  &-\frac{\Delta}{2} -\lambda\tau
\end{pmatrix}~.
\end{equation}

\begin{figure}[b]
  \centering
  \includegraphics[width=0.8\linewidth]{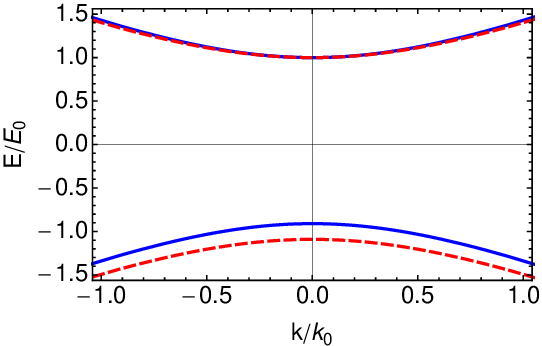}
\caption{Spin-resolved band dispersions for monolayer MoS$_2$ at the $\vek{K}$ point.
Spin-$\uparrow$($\downarrow$) bands are shown as the blue solid (red dashed) curves.
The unit scales for energy $E$ and wave vector $\vek{k}$ (measured from $\vek{K}$) are
given in terms of band-structure parameters as $E_{0}\equiv\Delta/2$ and $k_{0}\equiv
\Delta/(2at)$. Reversal of all spin labels yields the corresponding band dispersions at the
$\vek{K'}(\equiv -\vek{K})$ point.}
\label{fig:dispersion}
\end{figure}

The eigenenergies and eigenstates of the Hamiltonian $H_0^\tau$ are straightforwardly
obtained as
\begin{equation}
E^{(\tau,s)}_{\kk\alpha}=\frac{E_0}{2}\left(s \tau \bar\lambda + \alpha\sqrt{4 \bar k^2+
m_{s\tau}^2}\right)
\end{equation}
and
\begin{equation}
\psi^{(\tau,s)}_{\kk\alpha} = \frac{1}{\sqrt{2}}
\begin{pmatrix}
\alpha \left[1 + \frac{\alpha m_{s\tau}}{\sqrt{4\bar k^2 + m_{s\tau}^2}}\right]^{\frac{1}{2}}
\\[3mm]
\tau\left[1 - \frac{\alpha m_{s\tau}}{\sqrt{4\bar k^2 + m_{s\tau}^2}}\right]^{\frac{1}{2}}
\ee^{-i\tau\theta} 
\end{pmatrix}\otimes\ket{s}~,
\end{equation}
respectively, where $\alpha=1\,(-1)$ for conduction electrons (valence-band holes),
$s=\pm1$ labels the eigenstates of $s_z$, and $m_\pm\equiv2\mp\bar\lambda$. We
introduced dimensionless quantities $\bar k=k/k_0$, $\bar\lambda\equiv\lambda/E_0$,
with unit scales for energy and wave vector given by $E_{0}\equiv\Delta/2$ and
$k_{0}\equiv\Delta/(2at)$, respectively.~\footnote{In the limit $\lambda\to 0$, the
effective Hamiltonian (\ref{eq:MoS2Ham}) is equivalent to a Dirac model with effective
speed of light $c_{\text{eff}}\equiv a t/\hbar$ and effective rest mass $M_{\text{eff}}
\equiv \hbar^2 \Delta/(2 a^2 t^2)$. Hence , the scales $E_0$ and $k_0$ can be
associated with an effective rest energy $M_{\text{eff}} c_{\text{eff}}^2$ and inverse
Compton wave length $M_{\text{eff}} c_{\text{eff}}/\hbar$, respectively. Later on we
choose $\chi_0=2 k_0^2/(\pi E_0)\equiv 2 M_{\text{eff}}/(\pi\hbar^2)$ as the unit for
the spin-susceptibility tensor, as it corresponds to the density of states for a 2D
system of free electrons with effective mass $M_{\text{eff}}$ and four-fold flavor
degeneracy.} Figure~\ref{fig:dispersion} shows the electron and hole band dispersions
obtained for MoS$_2$. Note that SOC gives rise to an energy splitting for the hole
(valence-band) excitations that is finite ($\equiv 2\lambda$) even at the band edge.
For non-zero wave vector, inter-band coupling induces a spin splitting also for
conduction electrons, but its magnitude is suppressed because of the relatively
large band gap.

When the Fermi energy $E_{\text{F}}$ is above (below) the conduction-band
(valence-band) edge, the system is electron-doped (hole-doped). The Fermi
wave vectors for electronic excitations associated with the spin-split bands in
each valley are then given by $k^{(s\tau)}_{\text{F}}$, with
\begin{equation}\label{eq:Fermiwavevec}
k^{(\pm)}_{\text{F}} = k_0\, \sqrt{\left[(E_{\text{F}}/E_0) - 1 \right] \left[(E_{\text{F}}
/E_0) + 1 \mp \bar\lambda\right]} \quad .
\end{equation}
The total sheet density $n$ of charge carriers can be related to the Fermi wave
vectors via
\begin{equation}
n = \frac{1}{4\pi} \sum_{\tau, s} \left( k_{\text{F}}^{(s \tau)} \right)^2 \equiv 
\frac{g_v}{4\pi} \left[ \left( k_{\text{F}}^{(+)} \right)^2 + \left( k_{\text{F}}^{(-)}
\right)^2 \right] \, ,
\end{equation}
where $g_v=2$ is the degeneracy factor associated with the valley degree of
freedom. In the following, it will be useful to also define a density-related average
Fermi wave number $k_{\text{F}}$ such that $n = g_v g_s k_{\text{F}}^2/(4\pi)$,
with real-spin degeneracy factor $g_s=2$. Obviously we have
\begin{equation}
k_{\text{F}} = \left\{ \begin{array}{cl} k_0 \left[\left(\frac{E_{\text{F}}}{E_0}
\right)^2 - 1\right]^{\frac{1}{2}} & E_{\text{F}} > E_0 \mbox{ or } E_{\text{F}} < -E_0
-\lambda \\[3mm] k_{\text{F}}^{(+)} & -E_0 + \lambda > E_{\text{F}} > -E_0 -\lambda
\end{array}\right. .
\end{equation}

%%%%%%%%%%%%%%%%%%%%%%%%%%%%%%%%%%%%%%%%%%%%%%%
%%%%%%%%%%%%%%       SPIN SUSCEPTIBILITY     %%%%%%%%%%%%%%%%%
%%%%%%%%%%%%%%%%%%%%%%%%%%%%%%%%%%%%%%%%%%%%%%%
\subsection{Spin susceptibility for a multi-band system}

The influence of spin-dependent external stimuli on a many-particle system can be quite
generally discussed, within linear-response theory, in terms of the spin susceptibility
given by\cite{giu05}
\begin{eqnarray}\label{eq:SpinSusR}
\chi_{ij}(\rr-\rr')=-\frac{i}{\hbar}\int_0^\infty dt~\ee^{-\eta t}~\langle[S_i(\rr,t),S_j(\rr',0)]\rangle
\, .
\end{eqnarray}
Here $S_j(\rr)$ denotes a general Cartesian component of the spin-density operator
(we measure spin in units of $\frac{\hbar}{2}$), and $\rr$ is the position vector in the $xy$-plane.
We can express $S_j(\rr)$ in terms of the second-quantized particle creation and annihilation
operators $\Psi^\dagger$, $\Psi$ and the spin matrices $\hat{\mathcal J}_j$ as $S_j(\rr)=
\Psi^\dagger(\rr) \hat{\mathcal J}_j \Psi(\rr)$. As particle excitations are generally
superpositions of contributions from the individual valleys, we represent the particle
operator as a spinor, $\Psi(\rr)=\big(\Psi^{(+)}(\rr),\Psi^{(-)}(\rr)\big)$. In terms of energy
eigenstates and their annihilation operators $c^{(\tau,s)}_{\kk\alpha}$, the contributions
for each valley can be expressed as
\begin{eqnarray}
\Psi^\tau(\rr)=\sum_{s,\alpha} \int\frac{d^2k}{(2\pi)^2}~\ee^{i (\kk+ \tau{\bf K})\rr}
~\psi^{(\tau,s)}_{\kk\alpha}~ c^{(\tau,s)}_{\kk\alpha}~.
\end{eqnarray}
With these definitions, the spin susceptibility in Eq.~(\ref{eq:SpinSusR}) can be
written as the Fourier transform of the wave-vector-dependent spin susceptibility
$\chi_{ij}({\bf q})$,
\begin{equation}\label{eq:Fouriertrafo}
\chi_{ij}({\bf R})=\int\frac{d^2q}{(2\pi)^2}~\ee^{i{\bf q}{\bf R}}~\chi_{ij}({\bf q})~,
\end{equation}
where ${\bf R}\equiv\rr-\rr'$ and~\footnote{We only include contributions to the spin
susceptibility that involve intra-valley excitations, as has been done in a recent
calculation of the charge response\cite{sch13}. In principle, inter-valley terms
exist, but these are oscillating rapidly in real space~\cite{bre07} and are therefore
only relevant for physical observables on microscopic scales.}
\begin{subequations}
\begin{eqnarray}\label{eq:SpinSusq}
\chi_{ij}({\bf q})&=&\sum_{s,s',\tau}\sum_{\alpha,\beta}
\int\frac{d^2k}{(2\pi)^2}~\mathscr{W}_{ij(\kk,\kk+\qq,\alpha,\beta)}^{s,s',\tau}
\nonumber\\[2mm]&&\hspace{1.5cm}\times~\frac{n_{\text{F}}\big(
E^{(\tau,s)}_{\kk\alpha}\big) - n_{\text{F}}\big(E^{(\tau,s')}_{\kk+\qq \beta}\big)}
{E^{(\tau,s)}_{\kk\alpha}-E^{(\tau,s')}_{\kk+\qq\beta}+i\eta}~, \quad
\end{eqnarray}
with matrix elements
\begin{eqnarray}
\mathscr{W}_{ij(\kk,\kk',\alpha,\beta)}^{(s,s',\tau)}=
\left[(\psi^{(\tau,s)}_{\kk\alpha})^\dagger\hat J_i \psi^{(\tau,s')}_{\kk'\beta}\right]
\left[(\psi^{(\tau,s')}_{\kk'\beta})^\dagger\hat J_j \psi^{(\tau,s)}_{\kk\alpha}\right]~.
\nonumber\\
\end{eqnarray}
\end{subequations}
See the Appendix for more mathematical details. Here $n_{\text{F}}(\cdot)$
denotes the Fermi function. For our cases of interest, the two valleys make identical
contributions to the spin susceptibility, hence we can account for the valley degree of
freedom by a degeneracy factor $g_v=2$.

It follows from the structure of the spin matrices, Eqs.~(\ref{eq:SpinMat}), that 
the in-plane components $\chi_{xx}(\qq)=\chi_{yy}(\qq)$ contain contributions
only for $s\neq s'$. In contrast, $\chi_{zz}(\qq)$ has only terms with $s=s'$
contributing, thus $\chi_{zz}(\qq)$ is proportional to the Lindhard function
$\chi_0(\qq)$ calculated in Ref.~\onlinecite{sch13}. By similar arguments, it can
be established that all off-diagonal elements of the spin-susceptibility tensor vanish. 
As the Hamiltonian (\ref{eq:MoS2HamMat}) has axial symmetry, it follows that the
spin susceptibility depends only on the magnitude $q\equiv |\qq|$ of the wave vector
$\qq$.

%%%%%%%%%%%%%%%%%%%%%%%%%%%%%%%%%%%%%%%%%%%%%%%
%%%%%%%%%%%%%%%        ELECTRON DOPED      %%%%%%%%%%%%%%%%%
%%%%%%%%%%%%%%%%%%%%%%%%%%%%%%%%%%%%%%%%%%%%%%%
\section{Spin susceptibility of electrons: Extrinsic vs.\ intrinsic contributions\label{sec:susElec}}

\begin{figure*}[t]
\includegraphics[width=0.85\columnwidth]{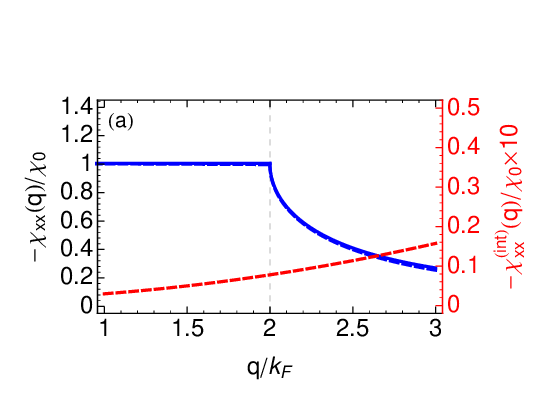}\hspace{1cm}
\includegraphics[width=0.85\columnwidth]{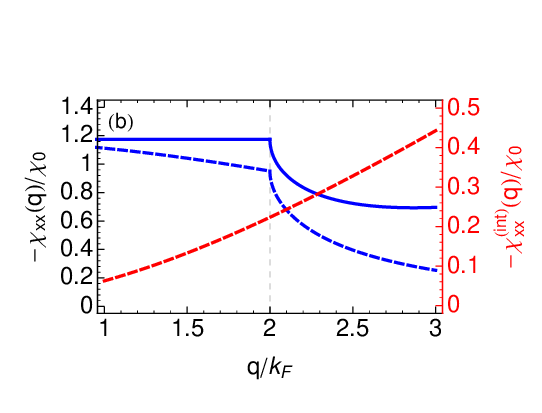}
\caption{\label{fig:electrondopedxx} 
In-plane component $\chi_{xx}(q)\equiv \chi_{xx}^{(\text{ext})}(q) + \chi^{(\text{int})}_{xx}(q)$
of the wave-vector-dependent spin-susceptibility tensor for electron-doped monolayer MoS$_2$
(blue solid curves). We also plot the extrinsic (intrinsic) contributions $\chi_{xx}^{(\text{ext})}$
$\big(\chi^{(\text{int})}_{xx}\big)$ separately as the blue (red) dashed curves.  Panel (a) [(b)]
is for electron density $n=1\times 10^{12}\,$cm${}^{-2}$ [$5\times10^{13}\,$cm${}^{-2}$].
Notice the sharp feature exhibited by $\chi_{xx}$ and $\chi_{xx}^{(\text{ext})}$ for $q\approx
2 k_{\text{F}}$ and the different ordinate scale for $\chi^{(\text{int})}_{xx}$.}
\end{figure*}

In this Section, we consider the situation where the Fermi energy is above the conduction-band
edge, i.e., $E_{\text{F}} > \Delta/2$. As in the previously considered case of the dielectric
polarizability of monolayer graphene~\cite{and06,hwa07,pya09,sch11,sch12}, the spin-response
function of the electron-doped system can be separated into an extrinsic contribution that is
entirely due to the occupied states in the conduction band and the intrinsic contribution arising
from the completely filled valence band. For the non-vanishing diagonal elements, we find
$\chi_{jj}(\qq)= \chi_{jj}^{(\text{ext})}(\qq) + \chi_{jj}^{(\text{int})}(\qq)$, with
\begin{subequations}\label{eq:ExInt}
\begin{eqnarray}
\chi_{jj}^{(\text{ext})}(\qq) &=& \sum_{s,s',\tau\atop \delta=\pm 1}
\int\frac{d^2k}{(2\pi)^2}~~n_{\text{F}}\big(E^{(\tau,s)}_{\kk +}\big) \nonumber \\
&& \hspace{-1.3cm} \times\left[
\frac{\mathscr{W}_{jj(\kk,\kk+\qq,+,+)}^{(s,s',\tau)}}{E^{(\tau,s)}_{\kk +}-E^{(\tau,s')}_{\kk
+\qq +} +i\eta\delta}+\frac{\mathscr{W}_{jj(\kk,\kk+\qq,+,-)}^{(s,s',\tau)} }{E^{(\tau,s)}_{\kk +}
-E^{(\tau,s')}_{\kk+\qq -}+i\eta\delta}\right] \, , \nonumber \\[2mm] \\[3mm]
\chi_{jj}^{(\text{int})}(\qq) &=&
-\sum_{s,s',\tau \atop \delta=\pm 1}
\int\frac{d^2k}{(2\pi)^2} ~~\frac{\mathscr{W}_{jj(\kk,\kk+\qq,+,-)}^{(s,s',\tau)}~ n_{\text{F}}
\big(E^{(\tau,s)}_{\kk -}\big)}{E^{(\tau,s)}_{\kk +} - E^{(\tau,s')}_{\kk+\qq -}
+i\eta\delta}~. \label{eq:Int}\nonumber \\
\end{eqnarray}
\end{subequations}
The expressions in Eqs.~(\ref{eq:ExInt}) have been obtained from Eq.~(\ref{eq:SpinSusq})
using the axial symmetry of the Hamiltonian (\ref{eq:MoS2HamMat}). In the
zero-temperature limit (which we employ in the following), the Fermi functions are
$n_{\text{F}}\big( E^{(\tau,s)}_{\kk +} \big)=\Theta\big(k^{(s\tau)}_{\text{F}}-k\big)$ and
$n_{\text{F}}\big( E^{(\tau,s)}_{\kk -} \big)=1$, respectively.

\subsection{In-plane spin-susceptibility component $\mathbf{\chi_{xx}}$}

An explicit calculation of the extrinsic contribution to the in-plane spin susceptibility
tensor element yields
\begin{widetext}
\begin{eqnarray}\label{eq:chixxext}
\chi^{\text{(ext)}}_{xx}(q)&=&-\frac{\chi_0}{4}\Biggl\{\frac{2\bar q^4+{\mathcal F}
\bar\lambda^2 (6+\bar\lambda+{\mathcal G})-\bar q^2[{\mathcal G}~{\mathcal F}+
\bar\lambda({\mathcal F}+8\bar\lambda)]}{(\bar q^2-\bar\lambda^2)^2}+2\left(\sqrt{4
\big({\bar k^{(+)}_{\text{F}}}\big)^2+m_+^2}-m_+\right) \Biggr.\nonumber \\[3mm]
&&\hspace{1.5cm}
\Biggl. +\frac{\left[\bar q^6-2\bar q^2\bar\lambda^2 +2\bar\lambda^4(\bar\lambda^2-4)
-\bar q^4(\bar\lambda^2+4)\right]}{(\bar\lambda^2-\bar q^2)^{5/2}}\ln{\frac{\bar q^2
\left(\sqrt{\bar\lambda^2-\bar q^2}-2\right)}{{\mathcal F}\sqrt{\bar\lambda^2-\bar q^2}+
\bar\lambda^2({\mathcal G}-m_+)-\bar q^2({\mathcal G}+\bar\lambda)}}\Biggr\}~,
\quad
\end{eqnarray}
with $\chi_0 = g_v g_s k_0^2/(2\pi E_0)\equiv\Delta/(\pi a^2 t^2)$. We have also used
the abbreviations
\begin{equation}
{\mathcal F} = \sqrt{\bar q^4-2\bar q^2\bar \lambda({\mathcal G}-m_+)-2\bar
\lambda^2 m_+({\mathcal G}-m_+)+4\kappa_q^2(\bar\lambda^2-\bar q^2)} \,\, ,
\end{equation}
\end{widetext}
\begin{eqnarray}
{\mathcal G} &=& \sqrt{4\kappa_q^2+m_+^2} \quad , \\
\kappa_q &=& \mathcal{K}_q \, \Theta \left( k_{\text{F}}^{(+)} + k_{\text{F}}^{(-)} - q
\right)+ \bar k_{\text{F}}^{(+)} \, \Theta \left( q - k_{\text{F}}^{(+)} - k_{\text{F}}^{(-)}
\right) , \nonumber \\ \\ \label{eq:kappa_q}
\mathcal{K}_q &=& \frac{\bar q^3-\bar q\bar\lambda(\bar\lambda-2)-\bar q\bar\lambda
\sqrt{4+\bar q^2-\bar\lambda^2}}{2(\bar q^2-\bar\lambda^2)} \quad .
\end{eqnarray}
In the limit $q\to 0$, the result
\begin{eqnarray}\label{eq:extZero}
\chi^{\text{(ext)}}_{xx}(0) &=& -\frac{\chi_0}{2}\Biggl(\frac{(4-\bar\lambda^2)~
\text{arctanh}\frac{\bar\lambda}{2}}{\bar\lambda}
\Biggr.\nonumber\\[2mm] {}&& \hspace{1.5cm}\Biggl.
+\sqrt{4\big({\bar k^{(+)}_{\text{F}}}\big)^2+m_+^2}-m_+\Biggl) \quad
\end{eqnarray}
is found.

\begin{figure*}[t]
\includegraphics[width=0.85\columnwidth]{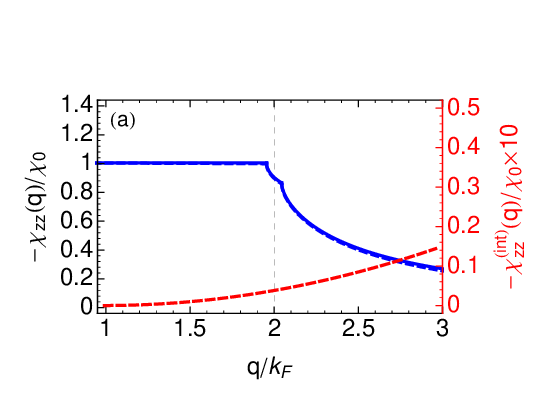}\hspace{1cm}
\includegraphics[width=0.85\columnwidth]{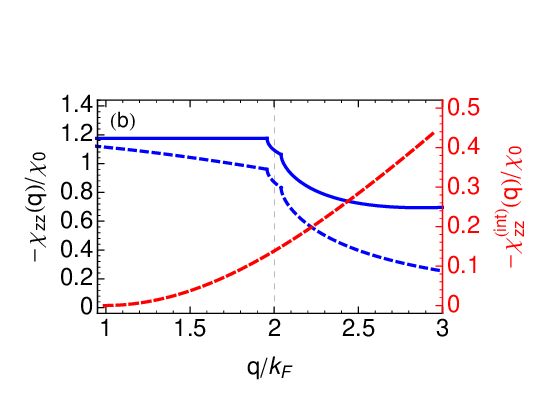}
\caption{\label{fig:electrondopedzz}
Spin-susceptibility component $\chi_{zz}(q)\equiv \chi_{zz}^{(\text{ext})}(q) +
\chi^{(\text{int})}_{zz}(q)$ for electron-doped monolayer MoS$_2$ (blue solid curves).
The extrinsic (intrinsic) contributions $\chi_{zz}^{(\text{ext})}$ $\big(\chi^{(\text{int})}_{zz}
\big)$ are also plotted as the blue (red) dashed curves.  Panel (a) [(b)] shows results for
electron density $n=1\times 10^{12}\,$cm${}^{-2}$ [$5\times10^{13}\,$cm${}^{-2}$]. Sharp
features are exhibited by $\chi_{zz}$ and $\chi_{zz}^{(\text{ext})}$ for $q=k_{\text{F}}^{(+)},
k_{\text{F}}^{(-)}$. Note the different scale for $\chi^{(\text{int})}_{zz}$.}
\end{figure*}

The intrinsic contribution can be expressed as
\begin{eqnarray}\label{eq:chixxint}
\chi^{\text{(int)}}_{xx}(q) &=& - \left. \chi^{\text{(ext)}}_{xx}(q)\right|_{\kappa_q \equiv
\mathcal{K}_q} + \chi^{\text{(ext)}}_{xx}(0)+\chi^{\text{(int)}}_{xx}(0)\,\, , \nonumber \\
\end{eqnarray}
with $\chi^{\text{(ext)}}_{xx}(0)$ from Eq.~(\ref{eq:extZero}). In contrast to the static
dielectric polarizability of both monolayer MoS$_2$\cite{sch13} and monolayer
graphene~\cite{and06,hwa07,pya09,sch11,sch12}, the in-plane spin susceptibility of
electron-doped transition metal dichalcogenides is found to have a finite intrinsic
contribution for $q\to 0$,
\begin{equation}\label{eq:chixxq0}
\chi^{\text{(int)}}_{xx}(0)=-\frac{\chi_0}{2}
\left(\frac{2\bar\lambda-(4-\bar\lambda^2)~
\text{arctanh}\frac{\bar\lambda}{2}}{\bar\lambda}\right) \quad .
\end{equation}
As the expression (\ref{eq:chixxq0}) vanishes for $\lambda\to 0$, the finite
$\chi^{\text{(int)}}_{xx}(0)$ is a SOC effect. Combining Eqs.~(\ref{eq:chixxq0}) and
(\ref{eq:extZero}) yields the $q\to 0$ limit of the in-plane spin-susceptibility tensor
component in the electron-doped case given by
\begin{subequations}\label{eq:sumxxq0}
\begin{eqnarray}
\chi_{xx}(0) &=& - \frac{\chi_0}{2}\left(\sqrt{4\big(\bar k^{(+)}_{\text{F}}\big)^2+m_+^2}+
\bar\lambda \right) \, , \\
&\equiv& -\chi_0\, \frac{E_{\text{F}}}{E_0} \quad .
\end{eqnarray}
\end{subequations}

Figure~\ref{fig:electrondopedxx} illustrates the behavior of $\chi_{xx}(q)$ and also shows
the individual extrinsic and intrinsic contributions. A cancellation of the latters' $q$ dependences
yield a constant $\chi_{xx}(q)$ for $q \le k^{(+)}_{\text{F}}+k^{(-)}_{\text{F}}$ ($\approx 2
k_{\text{F}}$ typically), which is followed by an abrupt decrease of the spin susceptibility for
$q > k^{(+)}_{\text{F}}+k^{(-)}_{\text{F}}$. The general line shape is
similar to the one found for an ordinary 2D electron gas~\cite{giu05} in the absence of spin-orbit
coupling, and the plateau behavior is also exhibited by response functions of monolayer
graphene~\cite{and06,hwa07,pya09,sch11}. The fact that only a single sharp feature appears
in $\chi_{xx}(q)$, even though there are two Fermi surfaces for the values of density used
for the plots, originates from the in-plane response being governed by transitions between
eigenstates with opposite $s_z$ quantum number. Furthermore, in contrast to the case of
an ordinary 2D electron gas, the plateau value of $\chi_{xx}(q)$ is density dependent [see
Eq.~(\ref{eq:sumxxq0})], but this dependence is much weaker than in the case of
graphene\cite{and06,hwa07,pya09,sch11} because of the relatively large value of the gap
parameter $\Delta$.

\subsection{Perpendicular spin-susceptibility component $\mathbf{\chi_{zz}}$}

The extrinsic contribution to the spin-susceptibility component describing the response in
the direction perpendicular to the 2D material's plane is the sum of terms arising from  the
individual spin-split bands,
\begin{subequations}
\begin{equation}\label{eq:chizzext}
\chi^{\text{(ext)}}_{zz}(q) = \sum_s \chi^{\text{(ext)}}_{zz, s}(q) \quad ,
\end{equation}
\begin{widetext}
\begin{eqnarray}\label{eq:chizzSpin}
\chi^{\text{(ext)}}_{zz, s}(q)&=&\frac{\chi_0}{8} \left\{m_s-2\sqrt{4\big(\bar
k^{(s)}_{\text{F}}\big)^2+m_s^2}-\frac{(m_s^2-\bar q^2)}{\bar q}\arctan\frac{\bar q}{m_s}\right\}
\Theta\big(2 k^{(s)}_{\text{F}}-q \big)\nonumber\\
&& \hspace{-0.8cm}+\,\frac{\chi_0}{8} \left\{m_s-2\sqrt{\big(\bar k^{(s)}_{\text{F}}
\big)^2+m_s^2}+ \frac{1}{\bar q} \sqrt{\left[4 \big(\bar k^{(s)}_{\text{F}}\big)^2+m_s^2\right]
\left[\bar q^2-4\big(\bar k^{(s)}_{\text{F}}\big)^2\right]}-\frac{m_s^2-\bar q^2}{\bar q}\Biggl[
2\arcsin\sqrt{\frac{4\big(\bar k^{(s)}_{\text{F}})^2+m_s^2}{m_s^2+\bar q^2}} \right.\nonumber\\
&&\hspace{-0.8cm}\left. -\, 2\arctan\frac{m_s}{\bar q}+\frac{1}{2}\arctan\left(\frac{m_s}{2\bar q}
-\frac{\bar q}{2m_s}\right)-\frac{1}{2}\arctan\frac{8\big(\bar k^{(s)}_{\text{F}}\big)^2+m_s^2-
\bar q^2}{2 \sqrt{4\big(\bar k^{(s)}_{\text{F}}\big)^2+m_s^2}\sqrt{\bar q^2-4\big(\bar
k^{(s)}_{\text{F}}\big)^2}}\Biggr]\right\}\Theta\big(q-2k^{(s)}_{\text{F}}\big) \, . \quad
\end{eqnarray}
\end{widetext}
\end{subequations}
In the $q\to 0$ limit, Eq.~(\ref{eq:chizzext}) yields
\begin{eqnarray}\label{eq:extzzq0}
\chi^{\text{(ext)}}_{zz}(0)=-\frac{\chi_0}{4}\sum_s\sqrt{4\big(\bar k^{(s)}_{\text{F}})^2
+m_s^2}~,
\end{eqnarray}
thus $-\chi^{\text{(ext)}}_{zz}(0)$ corresponds to the density of states at the Fermi energy.
For the intrinsic contribution, the expression
\begin{eqnarray}\label{eq:chizzint}
\chi^{\text{(int)}}_{zz}(q)=-\frac{\chi_0}{8}\sum_s
\left[ m_s-\frac{m_s^2-\bar q^2}{\bar q} \arctan\frac{\bar q}{m_s}\right]
\end{eqnarray}
is found, which vanishes in the limit $q\to 0$. As a result, $\chi_{zz}(0)\equiv
\chi^{\text{(ext)}}_{zz}(0)$, and we find using Eq.~(\ref{eq:extzzq0})
\begin{equation}\label{eq:plateauEq}
\chi_{zz}(0) = -\chi_0 \, \frac{E_{\text{F}}}{E_0} \equiv \chi_{xx}(0) \quad .
\end{equation}
 
The line shape of $\chi_{zz}(q)$ is shown in Fig.~\ref{fig:electrondopedzz} for band-structure
parameters of MoS$_2$ and two density values. As in the case of the in-plane spin-susceptibility
component, a cancellation of $q$ dependences from the extrinsic and intrinsic contributions results
in a plateau for $\chi_{zz}(q)$ for wave vectors smaller than a threshold value (here:
$2 k^{(+)}_{\text{F}}$). While the plateau value is the same as for $\chi_{xx}(q)$, its width is different.
Also in contrast to the behavior of $\chi_{xx}(q)$, two sharp features at $q=k^{(+)}_{\text{F}}$ and
at $q=k^{(-)}_{\text{F}}$ signify the existence of the two Fermi surfaces. However, the line shapes
of the in-plane and perpendicular spin-susceptibility components become very similar again for
$q > 2k^{(-)}_{\text{F}}$. Hence, except for wave vectors within the region close to the two Fermi
wave vectors, the spin response of charge carriers in electron-doped transition metal
dichalcogenides is isotropic and very similar to that of an ordinary 2D electron gas~\cite{giu05}.
Differences to the standard behavior will therefore occur in the oscillations of the spin
susceptibility in real space [Eq.~(\ref{eq:Fouriertrafo})] whose wave length and beating
pattern is governed by the sharp features in $\chi_{ij}(q)$.
 
%%%%%%%%%%%%%%%%%%%%%%%%%%%%%%%%%%%%%%%%%%%%%%%
%%%%%%%%%%%%%%%%         HOLE DOPED     %%%%%%%%%%%%%%%%%%%
%%%%%%%%%%%%%%%%%%%%%%%%%%%%%%%%%%%%%%%%%%%%%%%
\section{Spin susceptibility of holes: In-plane/out-of-plane anisotropy\label{sec:susHole}}

Specializing the general definition (\ref{eq:SpinSusq}) for the spin-susceptibility tensor to the
situation where only states in the valence band are occupied [i.e., for $n_{\text{F}}
\big(E^{(\tau,s)}_{\kk +}\big) \equiv 0$], we obtain
\begin{eqnarray}\label{eq:holesusc}
&&\chi_{jj}(\qq)=\sum_{s,s',\tau \atop \delta=\pm 1}
\int\frac{d^2k}{(2\pi)^2}~n_{\text{F}}\big(E^{(\tau,s)}_{\kk -}\big)\nonumber\\[2mm]
&&\times\left[\frac{\mathscr{W}_{jj(\kk,\kk+\qq,-,-)}^{(s,s',\tau)}}{E^{(\tau,s)}_{\kk -}-
E^{(\tau,s')}_{\kk+\qq -}+i\eta\delta}+\frac{\mathscr{W}_{jj(\kk,\kk+\qq,-,+)}^{(s,s',\tau)}}
{E^{(\tau,s)}_{\kk -}-E^{(\tau,s')}_{\kk+\qq +}+i\eta\delta}\right] . \quad
\end{eqnarray}
Note the analogy of the expression (\ref{eq:holesusc}) with that of the extrinsic part of the
electron-doped case [cf.\ Eq.~(\ref{eq:ExInt})].~\footnote{If we were to adopt the hole
picture by defining $\tilde n_{\text{F}} = 1 - n_{\text{F}}$ as the distribution function of
charge carriers, the expression (\ref{eq:holesusc}) could be written, in full analogy to the
electron-doped case, as the sum of the intrinsic contribution and an extrinsic part that
vanishes in the limit of zero hole density. While we have used the electron picture
throughout, the formulae given in our work make it possible to easily find the equivalent
results for the hole picture.}

\begin{figure*}[t]
\includegraphics[height=4.2cm]{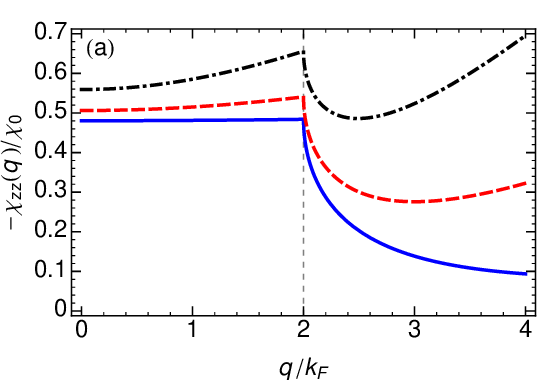}\hspace{1.5cm}
\includegraphics[height=4.2cm]{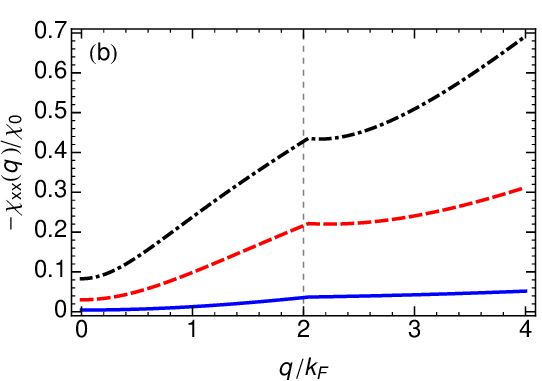}
\caption{\label{fig:hole_sus}%
Spin response of hole-doped monolayer MoS$_2$. Panel (a) [(b)] shows $\chi_{zz}(q)$
[$\chi_{xx}(q)$] obtained for a hole sheet density $n=1\times 10^{12}$cm${}^{-2}$ (blue solid
curve), $n=1\times 10^{13}$cm${}^{-2}$ (red dashed curve), and $n=3\times10^{13}$cm${}^{-2}$
(black dot-dashed curve). Noteworthy features are the strong in-plane/out-of-plane anisotropy of
the spin response, the deviations from ideal 2D-electron-gas behavior even for $\chi_{zz}(q)$,
which are getting more pronounced as the density increases, and the nonanalyticity exhibited
by $\chi_{zz}(q)$ [$\chi_{xx}(q)$] for $q = 2 k_{\text{F}}\equiv 2 k_{\text{F}}^{(+)}$ [$q = k_1>
2 k_{\text{F}}^{(+)}$].}
\end{figure*}

In  the following, we consider the situation where hole densities $n$ are small enough so
that only the upper-most of the two spin-split valence bands has empty states. This implies
$\Delta/2 - \lambda < -E_{\text{F}} < \Delta/2+\lambda $, and there will be only one Fermi surface
with radius $k_{\text{F}}^{(+)}\equiv\sqrt{4\pi n/g_v}$. For this situation, we obtain in the
zero-temperature limit the in-plane component of the spin-susceptibility tensor as
$\chi_{xx}(\qq)=\chi^{(1)}_{xx}(\qq)+\chi^{(2)}_{xx}(\qq)$, where
\begin{widetext}
\begin{eqnarray}\label{eq:chixxextholes}
&&\chi^{(1)}_{xx}(q)= \nonumber \\[3mm]
&& -\frac{\chi_0}{8}\Biggl\{\frac{2\bar q^4(1-\bar\lambda)+\bar\lambda^2\left[{\mathcal H}
(\tilde{\mathcal G}-6-\bar\lambda)+4\bar\lambda(4-\bar\lambda^2)\right]+\bar q^2[
\bar\lambda(8+{\mathcal H}-8\bar\lambda+6\bar\lambda^2)-\tilde{\mathcal G}
\,{\mathcal H}]}{(\bar q^2-\bar\lambda^2)^2}+2\left(\sqrt{4\big({\bar k^{(+)}_{\text{F}}}
\big)^2 +m_+^2}-m_+\right)
\Biggr.\nonumber \\[3mm] && \hspace{1.5cm} \Biggl.
+\,\frac{\left[\bar q^6-2\bar q^2\bar\lambda^2 +2\bar\lambda^4(\bar\lambda^2-4)-\bar q^4
(\bar\lambda^2+4)\right]} {(\bar\lambda^2-\bar q^2)^{5/2}}~\ln{\frac{\bar q^2
\left(2\bar\lambda-2+\sqrt{\bar\lambda^2-\bar q^2}\right)+2\bar\lambda m_+\left(\sqrt{\bar
\lambda^2-\bar q^2}+\bar\lambda\right)}{{\mathcal H}\sqrt{\bar\lambda^2-\bar q^2}+
\bar\lambda^2(m_+ + \tilde{\mathcal G})+\bar q^2(\bar\lambda-\tilde{\mathcal G})}}\Biggr\}~,
\end{eqnarray}
and $\chi^{(2)}_{xx}(\qq)=\chi^{(\text{int})}_{xx}(\qq)$ given in Eq.~(\ref{eq:chixxint}).
We have again introduced abbreviations
\begin{equation}
{\mathcal H} = \sqrt{\bar q^4+2\bar q^2\bar \lambda(\tilde{\mathcal G}+m_+)+2\bar\lambda^2
m_+(\tilde{\mathcal G}+m_+)+4\tilde\kappa_q^2(\bar\lambda^2-\bar q^2)} \quad ,
\end{equation}
\end{widetext}
\begin{eqnarray}
\tilde{\mathcal G} &=& \sqrt{4 \tilde\kappa_q^2+m_+^2} \quad , \\ \label{eq:kappaqhole}
\tilde\kappa_q &=& \mathcal{K}_q \, \Theta \left( k_1 - q \right) + \bar k_{\text{F}}^{(+)} \, \Theta
\left( q - k_1 \right) , \\
\frac{k_1}{k_0} &=& \bar k_{\text{F}}^{(+)} + \left[ \big(\bar k_{\text{F}}^{(+)}\big)^2+\bar\lambda
\left(\sqrt{4\big(\bar k_{\text{F}}^{(+)}\big)^2+m_+^2}-m_+\right)\right]^{\frac{1}{2}} , \nonumber \\
\end{eqnarray}
with ${\mathcal K}_q$ from Eq.~(\ref{eq:kappa_q}). Note that $\chi_{xx}(q)$ is nonanalytic at
$q=k_1 \,\big( > 2 k^{(+)}_{\text{F}}\big)$. In the limit $q \to 0$, Eq.~(\ref{eq:chixxextholes})
yields
\begin{eqnarray}\label{eq:chixxholeq0}
\chi_{xx}^{(1)}(0) =-\frac{\chi_0}{4}\left( \sqrt{4 \big({\bar k^{(+)}_{\text{F}}}\big)^2+m_+^2}-m_+\right)~.
\end{eqnarray}

The general result for the spin-susceptibility tensor component perpendicular to the plane is
obtained as
\begin{equation}
\chi_{zz}(q) = \sum_s \left(\chi_{zz, s}^{(1)}(q)~\Theta\big( k_{\text{F}}^{(s)} \big)+\chi_{zz, s}^{(2)}(q)\right)~,
\end{equation}
where $\chi_{zz, s}^{(1)}(q)=\chi_{zz, s}^{(\text{ext})}(q)$ and $\chi_{zz, s}^{(2)}(q)=
\chi_{zz, s}^{(\text{int})}(q)$, with the expressions for $\chi_{zz, s}^{(\text{ext})}(q)$ and
$\chi_{zz, s}^{(\text{int})}(q)$ from Eqs.~(\ref{eq:chizzSpin}) and (\ref{eq:chizzint}), respectively. 
Unlike in the electron-doped case, SOC does not give rise to the existence of two Fermi surfaces
for all hole densities. For our case of interest where hole densities are low enough such that
only the upper-most valence band has empty states, only a single Fermi surface exists.
In this situation, the $q\to 0$ limit yields
\begin{eqnarray}\label{eq:chizzholeq0}
\chi_{zz}(0)=-\frac{\chi_0}{4} \, \sqrt{4\big(\bar k^{(+)}_{\text{F}}\big)^2+m_+^2} \equiv \frac{\chi_0}{2}
\, \left( \frac{E_{\text{F}}}{E_0} - \frac{\bar\lambda}{2} \right)~, \nonumber \\
\end{eqnarray}
which corresponds to the density of states in the upper-most valence band. In contrast to the
in-plane spin-susceptibility component, $\chi_{zz}(q)$ is non-analytic at $q=2 k^{(+)}_{\text{F}}$.
Also, $|\chi_{zz}(0)| \gg |\chi_{xx}(0)|$ for typical hole densities, signifying a strong anisotropy of the
spin response.

The behavior of the spin response in the hole-doped case differs markedly from the electron-doped
situation. See Fig.~\ref{fig:hole_sus} for an illustration. 
As a first observation, a strong dependence
on hole-sheet density is apparent. In the low-density regime, the in-plane spin response is almost
uniformly very small, whereas $\chi_{zz}(q)$ has the line shape associated with the response functions
of an ordinary 2D electron system~\cite{giu05}. As the hole density increases, a pronounced peak
develops in $\chi_{xx}(q)$ for $q = k_1 \gtrsim 2 k_{\text{F}}^{(+)}$, and the plateau behavior of
$\chi_{zz}(q)$ disappears. Some of these features are very similar to those exhibited by the
spin response of 2D hole systems~\cite{ker13} realized by a quantum-well confinement in typical
semiconductor heterostructures~\cite{win03}. 
The nonanalyticity at (near) $q=2 k_{\text{F}}$
in $\chi_{zz}$ ($\chi_{xx}$) as well as the power-law behavior in its vicinity~\cite{she13}
determine the decay of the corresponding spin-susceptibility oscillations in real space. This and
other consequences of the unusual spin-response properties in the hole-doped case will be
discussed in greater detail in the following Section.

%%%%%%%%%%%%%%%%%%%%%%%%%%%%%%%%%%%%%%%%%%%%%%%
%%%%%%%%%%%%%%%%         OBSERVABLES     %%%%%%%%%%%%%%%%%%
%%%%%%%%%%%%%%%%%%%%%%%%%%%%%%%%%%%%%%%%%%%%%%%
\section{Physical consequences of unusual spin response in the hole-doped case\label{sec:obs}}

Based on the results presented in the previous Section, we consider spin-related physical
quantities for hole-doped monolayers of transition metal dichalcogenides. We start by discussing
the properties of hole-carrier-mediated exchange interaction between localized impurity spins.
Then the paramagnetic response of our system of interest is investigated. These examples serve
to illustrate the very different behavior of hole-doped systems, in contrast to the electron-doped
case that mirrors the properties of ordinary 2D electron gases.

\subsection{RKKY interaction and mean-field magnetism}

\begin{figure*}[t]
\includegraphics[height=4.2cm]{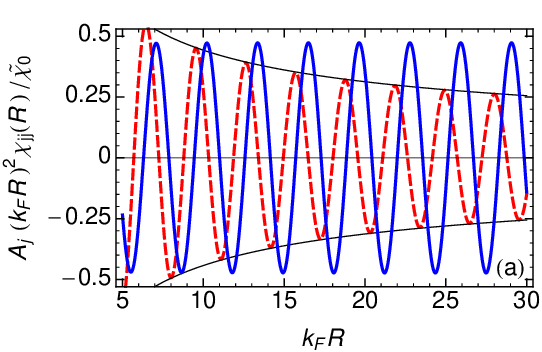}\hspace{1.5cm}
\includegraphics[height=4.2cm]{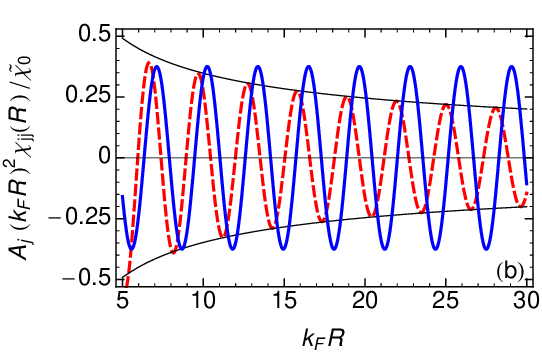}
\caption{Spin response functions in real space. To highlight deviations from the ordinary
2D-electron-gas behavior~\cite{giu05}, we plot the quantities $A_j \big(k_{\text{F}} R \big)^2
\chi_{jj}(R)$ in Panel (a) [(b)] obtained for a hole density $n=1\times 10^{12}$cm${}^{-2}$
[$n=3\times 10^{13}$cm${}^{-2}$]. The red dashed (blue solid) curve shows $\chi_{xx}$
($\chi_{zz}$). The decrease of the oscillation amplitude for $\big(k_{\text{F}} R \big)^2
\chi_{xx}(R)$ shows that the in-plane spin response decays faster than the $R^{-2}$
power law expected for ordinary 2D electron systems. The scale factors are
$A_x = 50$ [$5$] and $A_z = 1$ [$1$] in Panel (a) [(b)], and $\tilde\chi_0
=\chi_0 k_{\text{F}}^2/(2\pi) \equiv 2 n k_0^2/(\pi E_0)$.                                                 
\label{fig:FriedelOsc}}
\end{figure*}

We consider two localized impurity spins $\vek{I}^{(1)}$ and $\vek{I}^{(2)}$ that couple via a
contact interaction of strength $J$ to the local spin density of holes in a monolayer transition
metal dichalcogenide sample. In second-order perturbation theory, such a coupling gives rise
to an effective exchange interaction between the impurity-spin components that is described
by the RKKY Hamiltonian~\cite{rud54,kas56,yos57}  
\begin{equation}  
H_{\text{RKKY}}^{(1,2)}=-J^2  \sum_{i, j} ~ I^{(1)}_i \, I^{(2)}_j~\chi_{ij}(\vek{R})~.
\end{equation}
Here ${\mathbf R}$ is the distance vector between the locations of the two impurity spins,
and $\chi_{ij}(\vek{R})$ denotes the spin susceptibility in real space given by
Eq.~(\ref{eq:Fouriertrafo}). For our cases of interest, the spin susceptibility turns out to be
isotropic in its dependence on real-space position; $\chi_{ij}(\vek{R})\equiv \chi_{ij}(R)$.

Figure~\ref{fig:FriedelOsc} shows plots of the quantity $\big( k_{\text{F}} R\big)^2 \chi_{jj}
(R)$ for two realistic values of the hole density.\footnote{In the calculation of the Fourier
transform we introduce a smooth momentum cutoff factor $\text{exp}(-q/\Lambda)$
under the integral, with $\Lambda=\pi/a$, where $a$ is the lattice constant. See, e.g.,
S. Saremi, Phys. Rev. B \textbf{76}, 184430 (2007).}
The fact that $\chi_{zz}(R)\propto R^{-2}$
is clearly indicated by the constancy of the oscillations exhibited by the blue solid curves.
In contrast, the in-plane response function is seen to decay faster with distance $R$. Closer
inspection reveals that $\chi_{zz}(R)\propto R^{-5/2}$, i.e., shows behavior that
deviates from the expected $R^{-2}$ power law of a 2D Fermi
liquid.~\footnote{Previously, deviations from the $R^{-d}$ power-law decay of the
RKKY range function in a $d$-dimensional electron system have been shown to emerge as
the result of strong electron-electron correlations~\cite{she13} or for particular placements
of magnetic impurity atoms with respect to a material's crystal lattice~\cite{kir08,uch11}.
In contrast, the shorter-than-normal range of the in-plane spin response for MoS$_2$ found
here is exhibited by a uniform and non-interacting electron system.} Thus the in-plane
RKKY range function for monolayer transition metal dichalcogenides shows behavior that
is intermediate between that of an ordinary 2D electron gas~\cite{giu05} (or doped
graphene~\cite{bre07}) and undoped graphene~\cite{wun06}.

\begin{figure}[b]
\centering
\includegraphics[width=0.8\columnwidth]{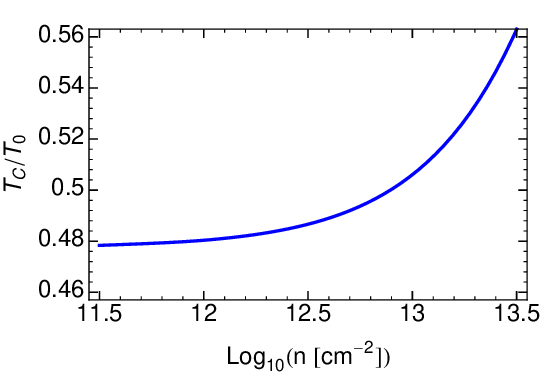}
\caption{Density dependence of the Curie temperature for hole-mediated easy-axis Ising
ferromagnetism of impurity spins in monolayer MoS$_2$. \label{fig:Curietemp}}
\end{figure}

In the low-density regime, the amplitude of $\chi_{zz}(R)$ can be more than an order of
magnitude larger than that of $\chi_{xx}(R)$; see Fig.~\ref{fig:FriedelOsc}(a). However,
as the density is increased, $\chi_{xx}(R)$ becomes appreciable and even reaches the
same magnitude as $\chi_{zz}(R)$; see Fig.~\ref{fig:FriedelOsc}(b). From the figure, it
is also apparent that the oscillations of $\chi_{xx}(R)$ and $\chi_{zz}(R)$ have a relative
phase shift that varies somewhat with $R$ and sometimes turns out to be close to $\pi/2$.
It follows from this observation that the lowest-energy state of two RKKY-coupled impurity
spins can change from the typically expected easy-axis configuration (both impurity spins
align in the direction perpendicular to the monolayer plane) to an easy-plane alignment
if the distance $R$ corresponds to a point where $\chi_{xx}(R)$ [$\chi_{zz}(R)$] has a
maximum [a zero].

Considering now a large number of impurity spins distributed, on average, homogeneously
with density $n_I$ in the material, the RKKY spin Hamiltonian can be treated using standard
mean-field theory~\cite{yos96}. For the hole-doped situation, we have $|\chi_{zz}(q)|_{q=0}
\gg |\chi_{xx}(q)|_{q=0}$, hence the spin system will exhibit Ising-type ferromagnetism with
Curie temperature given by 
\begin{subequations}
\begin{equation}\label{eq:Curietemp}
T_{\text{C}}=T_0 ~ \left. \frac{|\chi_{zz}(q)|}{\chi_0}\right|_{q=0}~,
\end{equation}
with the temperature scale
\begin{equation}
T_0=\frac{I(I+1)}{3}\, \frac{J^2}{k_{\text{B}}}\, n_{I}\, \frac{\Delta}{\pi a^2t^2}~.
\end{equation}
\end{subequations}
Here $I$ and $n_I$ denote the impurity spin quantum number and areal density of
impurities, respectively.
Due to the density dependence of $\chi_{zz}(q=0)$ [see Eq.~(\ref{eq:chizzholeq0})], the
Curie temperature can, in principle, be manipulated by the magnitude of hole doping.
However, as illustrated in Fig.~\ref{fig:Curietemp}, the range of realistic values for the hole
density allows for an adjustment of $T_{\text{C}}$ by only upto 10\% in the high-density
regime.

In general, mean-field predictions for transition temperatures are only a crude
approximation to reality, as the excitation of spin waves generally suppresses -- in some
cases, even destroys -- magnetic order.\cite{sim08} 

\subsection{Pauli paramagnetism and effective $g$-factor}

An external magnetic field generally couples to the hole carriers' spin via a Zeeman term
$H_{\text{Z}}=\kappa\, \mu_{\text{B}}\, B_j \hat J_j$, where $\mu_{\text{B}}$ is the Bohr
magneton and $2 \kappa$ the bulk valence-band $g$ factor.~\footnote{We adopt a notation
that is commonly used in semiconductor physics. See, e.g., Ref.~\onlinecite{suz74}.}
In the limit of a small
magnetic field, the paramagnetic susceptibility is given by
\begin{equation}\label{eq:paramagchi}
\chi_{\text{P},j}=(\kappa\,\mu_{\text{B}})^2~\chi^{(1)}_{jj}(q)|_{q=0}~,
\end{equation}
where $\chi^{(1)}_{jj}(q)$ are the spin susceptibilities of the hole-doped system for the in-plane
and out-of-plane response whose $q\to 0$ limits are shown in Eqs.~(\ref{eq:chixxholeq0})
and (\ref{eq:chizzholeq0}). It is possible to define a collective $g$-factor for the charge
carriers by expressing the paramagnetic susceptibility in terms of the density of states,
which is the zero-$q$ limit of the Lindhard function~\cite{giu05} $\chi_{\text{L}}(q)$, as
$\chi_{\text{P},j}=( g_j \mu_{\text{B}})^2\, \chi_{\text{L}}(0)/4$, and equate this with the
expression in Eq.~(\ref{eq:paramagchi}) to yield~\cite{ker13}
\begin{equation}\label{eq:gfactor}
g_j=2 \kappa \, \sqrt{\left. \frac{\chi^{(1)}_{jj}(q)}{\chi_{\text{L}}(q)}\right|_{q=0}} \quad .
\end{equation}

\begin{figure}[t]
  \centering
  \includegraphics[width=0.8\columnwidth]{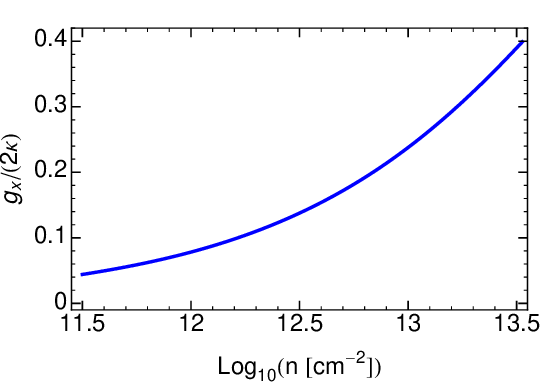}
\caption{Density dependence of the effective in-plane $g$-factor for hole-doped MoS$_2$.
\label{fig:effgfactor}}
\end{figure}

Our result for the out-of-plane spin response of holes in monolayer transition metal
dichalcogenides implies $g_z \equiv 2\kappa$, as $\chi_{zz}(q=0)$ turns out to be
equal to the density of states at the Fermi energy. However, the in-plane $g$-factor
shows unusual behavior, which is illustrated in Fig.~\ref{fig:effgfactor}. For small densities
(small $k_{\text{F}}$), $g_x$ is negligible [see Eq.~(\ref{eq:chixxholeq0})]. In contrast,
for large hole densities, $g_x$ can become of the same order of magnitude as $g_z$.

%%%%%%%%%%%%%%%%%%%%%%%%%%%%%%%%%%%%%%%%%%%%%%%
%%%%%%%%%%%%%%%%        SUMMARY    %%%%%%%%%%%%%%%%%%%%%
%%%%%%%%%%%%%%%%%%%%%%%%%%%%%%%%%%%%%%%%%%%%%%%
\section{Summary and Conclusions\label{sec:sum}}

We have calculated, and obtained analytical expressions for, the wave-vector-dependent static
spin susceptibility of charge carriers in monolayer transition metal dichalocogenides. Our
approach is based on the effective-mass model description of electronic excitations in these
materials. Very different behavior emerges for the cases of electron-doped and hole-doped
systems. We illustrate our findings using band parameters of MoS$_2$ monolayers.

Features exhibited for electron doping are similar, but not entirely analogous, to those associated
with ordinary 2D electron gases. The finite SOC results in deviations from the canonical line shape
near $q= 2 k_{\text{F}}$. Also, unlike the ordinary 2D electron system, the plateau value of the
spin response at small $q$ depends on the electron sheet density. See Eq.~(\ref{eq:plateauEq}).
The total response of the electron-doped system is obtained as the sum of an extrinsic part that
vanishes without the doping and an intrinsic contribution due to the completely filled valence band.

The hole-doped system shows marked deviations from the behavior expected from an ordinary
2D electron gas. In that, it mirrors some of the features of confined valence-band states in
semiconductor heterostructures~\cite{die97,ker13}. In particular, a strong anisotropy of the spin
susceptibility is exhibited, with the out-of-plane response being much stronger than the in-plane
response in the low-density limit. However, the in-plane response is enhanced as the hole
density increases and shows pronounced nonanalytic behavior near $q=2 k_{\text{F}}$. We
have investigated implications for spin-related physical quantities arising from the unusual
spin response of the hole-doped system. We show that the oscillations of the in-plane spin
response in real space decay faster than the typical $R^{-2}$ law that is expected for a 2D
Fermi liquid. Both the Curie temperature for hole-mediated easy-axis ferromagnetism and
the $g$ factor characterizing the Zeeman spin splitting due to an in-plane magnetic field
are found to be tunable by changing the hole density.

In this work, we have neglected effects due to disorder and electron-electron interactions,
which are known to, in principle, alter the spin response of ordinary 2D electron
systems~\cite{pal09}. Parameterization in terms of local field factors~\cite{giu05,pal09}
could be used to shed further light on how interactions renormalize the spin susceptibility
of monolayer transition metal dichalcogenides. As far as disorder is concerned, it can be
expected that important corrections to our results obtained in the clean limit will only arise
for low-enough carrier densities when the difference between $E_{\text{F}}$ and the band
edge is comparable in magnitude to the disorder-induced lifetime broadening~\cite{cap02,pal05}.
The latter turns out to be of the order of $\sim\,$0.01$\,$eV in typical samples~\cite{rad11} and,
therefore, is at least an order of magnitude smaller than all other relevant energy scales.

Our work adds to the understanding of monolayer transition metal dichalcogenides as a
new materials system whose charge carriers show behavior that is sometimes reminiscent
of -- but generally distinct from -- other 2D systems. The very different properties exhibited
by the electron-doped and hole-doped cases create the possibility for a versatile engineering
of electronic systems with specially tailored spin response.
To verify our theoretical results, electronic-transport experiments could be used to measure
the carrier spin susceptibility in the $q\to 0$ limit~\cite{zhu03,vak04}. Furthermore, monolayer
transition metal dichalcogenides would lend themselves as ideal samples for implementing
a recent proposal~\cite{stan13} for determining the full spatial structure of the spin susceptibility.

\begin{acknowledgments}
The authors thank R.~Asgari for numerous illuminating discussions.
\end{acknowledgments}

%%%%%%%%%%%%%%%%%%%%%%%%%%%%%%%%%%%%%%%%%%%%%%
%%%%%%%%%%%%%%%%        APPENDIX    %%%%%%%%%%%%%%%%%%%%%
%%%%%%%%%%%%%%%%%%%%%%%%%%%%%%%%%%%%%%%%%%%%%%%
\appendix*

\begin{widetext}

\section{Lehmann-type representation for the static spin susceptibility}

Here we give a short derivation of the expression for the wave-vector-dependent spin susceptibility
used in our work [Eq.~(\ref{eq:SpinSusq})] starting from the general formula (\ref{eq:SpinSusR}).
To this end, we explicitly write the spin operator as
\begin{equation}
S_i(\rr)=\sum_{\alpha,\beta}\int\frac{d^2k}{(2\pi)^2}\int\frac{d^2k'}{(2\pi)^2}~\ee^{i(\kk'-\kk)\rr}~(\psi_{\kk\alpha}^\dagger\hat J_i~\psi_{\kk'\beta})~c_{\kk\alpha}^\dagger c_{\kk'\beta}~,
\end{equation}
where we have used a greek index to include the quantum numbers for sublattice, spin and valley.
The time-dependent spin operator is given by
$S_i(\rr,t)=\ee^{\frac{i}{\hbar}H_0t}~S_i(\rr)~\ee^{-\frac{i}{\hbar}H_0t}=\ee^{i(E_{\kk\alpha}-E_{\kk'\beta})\frac{t}{\hbar}}S_i(\rr)$.
The commutator under the integral in (\ref{eq:SpinSusR}) is then given by
\begin{eqnarray}\label{eq:CommSpin}
[S_i(\rr,t),S_j(\rr')]&=&\sum_{\alpha,\beta,\gamma,\delta}
\int\frac{d^2k}{(2\pi)^2}\int\frac{d^2k'}{(2\pi)^2}\int\frac{d^2k''}{(2\pi)^2}\int\frac{d^2k'''}{(2\pi)^2}~\ee^{i(\kk'-\kk)\rr}\ee^{i(\kk'''-\kk'')\rr'}
\ee^{i(E_{\kk\alpha}-E_{\kk'\beta})\frac{t}{\hbar}}\nonumber\\[2mm]
&&{}(\psi_{\kk\alpha}^\dagger\hat J_i~\psi_{\kk'\beta})(\psi_{\kk''\gamma}^\dagger\hat J_j~\psi_{\kk'''\delta})~[c_{\kk\alpha}^\dagger c_{\kk'\beta},c_{\kk''\gamma}^\dagger c_{\kk'''\delta}]~.
\end{eqnarray}
The commutator in (\ref{eq:CommSpin}) involving the creation and annihilation operators is evaluated to give
$[~,~]=(2\pi)^2\delta^{(2)}(\kk'-\kk'')\delta_{\beta\gamma}c_{\kk\alpha}^\dagger c_{\kk'''\delta}-(2\pi)^2\delta^{(2)}(\kk-\kk''')
\delta_{\alpha\delta}c_{\kk''\gamma}^\dagger c_{\kk'\beta}$.
The equilibrium average of the same commutator gives $\langle[~,~]\rangle=(2\pi)^4\delta_{\beta\gamma}\delta_{\alpha\delta}\delta^{(2)}(\kk'-\kk'')\delta^{(2)}(\kk-\kk''')
[n_{\text{F}}(E_{\kk\alpha})-n_{\text{F}}(E_{\kk'\beta})]$.
Performing the summations and trivial integrations yields
\begin{eqnarray}\label{eq:CommSpin2}
\langle[S_i(\rr,t),S_j(\rr')]\rangle=\sum_{\alpha,\beta}
\int\frac{d^2k}{(2\pi)^2}\int\frac{d^2k'}{(2\pi)^2}~\ee^{i(\kk'-\kk)(\rr-\rr')}
\ee^{i(E_{\kk\alpha}-E_{\kk'\beta})\frac{t}{\hbar}}
(\psi_{\kk\alpha}^\dagger\hat J_i~\psi_{\kk'\beta})(\psi_{\kk'\beta}^\dagger\hat J_j~\psi_{\kk\alpha})~
[n_{\text{F}}(E_{\kk\alpha})-n_{\text{F}}(E_{\kk'\beta})].\nonumber\\
\end{eqnarray}
\end{widetext}
Making in (\ref{eq:CommSpin2}) the variable transformations $\kk'=\qq+\kk$ and ${\bf R}= \rr-\rr'$, and finally performing the
time integration on the r.h.s. of (\ref{eq:SpinSusR}) yields (\ref{eq:Fouriertrafo})-(\ref{eq:SpinSusq}) exactly.

%----------------------------------------------------------------------------------------------------------------------------------------------
%-------------------------------------------  Bibliography ------------------------------------------------------------
%----------------------------------------------------------------------------------------------------------------------------------------------
%
%\bibliography{MoS2susc}
%
%merlin.mbs apsrev4-1.bst 2010-07-25 4.21a (PWD, AO, DPC) hacked
%Control: key (0)
%Control: author (72) initials jnrlst
%Control: editor formatted (1) identically to author
%Control: production of article title (-1) disabled
%Control: page (0) single
%Control: year (1) truncated
%Control: production of eprint (0) enabled
%

\end{document}